# Instrument front-ends at Fermilab during Run II[*]

**Thomas Meyer**[a], **David Slimmer**[a] **and Duane Voy**[a]

[a] *Fermi National Accelerator Laboratory,*
*Batavia, IL, U.S.A.*

*E-mail*: tsmeyer@fnal.gov

ABSTRACT: The optimization of an accelerator relies on the ability to monitor the behavior of the beam in an intelligent and timely fashion. The use of processor-driven front-ends allowed for the deployment of smart systems in the field for improved data collection and analysis during Run II. This paper describes the implementation of the two main systems used: National Instruments LabVIEW running on PCs, and WindRiver's VxWorks real-time operating system running in a VME crate processor.

[*] Work supported by Fermi Research Alliance, LLC under Contract No. DE-AC02-07CH11359 with the United States Department of Energy.

**Contents**

**1. Introduction**

Instrumentation systems at Fermilab accelerators generally consist of a front-end which, in addition to data acquisition and processing, supports communication with the Accelerator Controls Network (ACNet) control system [1-3]. ACNet coordinates data transmission between the remote front-ends and ultimately the operators and users. The majority of instrumentation front-ends are based on three platforms: CAMAC, which is a lower performance system developed in the 1970's using proprietary hardware and software; LabVIEW™ [4], a National Instruments product that runs on a PC and has support for various data acquisition products; and VxWorks® [5], a real-time operation system from Wind River which at Fermilab is typically run in a VME crate processor. The instrumentation department within the accelerator division at Fermilab predominantly uses the latter two systems, and as such, they are the focus of this paper.

    Using two separate systems and platforms has allowed for the flexibility required for development of the varied systems needed by the operators and accelerator physicists. LabVIEW allows rapid prototyping and development of more complex systems. These systems often include motor and power supply control as well as data collection and manipulation. The LabVIEW system is a proprietary programming tool developed and licensed exclusively by National Instruments. It uses a graphical development interface to allow developers to create data flow diagrams. These diagrams, called VIs, often look like printed circuit boards where each IC represents another level of flow control and calculation. These separate circuits, or SubVIs, can contain their own text and graphically based diagnostics. This gives the developer the ability to create very complex tools and systems for both development and debugging of accelerator instrumentation systems.

    VxWorks, running on a VME platform and wide industry use, allows the Instrumentation Department to develop complex systems that need realtime data flow performance over ACNet.

This hardware/software platform has become a standard across the Accelerator complex allowing for easy integration into the ACNet control system.

VxWorks was originally brought into the Fermilab accelerator arena by power supply developers to control large bus power supplies which needed a fast realtime environment to control the Tevatron's bus current regulation. It was quickly adopted by the Accelerator Division's Controls Department for use in both device control and readback.

Built on the VME platform, VxWorks front ends have a very large industry support base to provide hardware and drivers for many uses. VME, being an open system, also allows for in-house development of hardware. In most of the accelerators' operational devices, Consumer Off The Shelf (COTS) boards are used along with Fermilab-developed hardware. This flexibility has allowed us to find unique solutions to the accelerators' instrumentation needs.

**Table 1.** LabVIEW and VxWorks front-ends used in the various accelerators including Main Injector (MI) and Tevatron (TeV).

| LabVIEW Front-Ends | VxWorks on VME Front-Ends |
| --- | --- |
| Flying Wires (TeV and MI) | Beam Position Monitors (TeV and MI) |
| Beam Line Tuners (TeV and MI) | Beam Loss Monitors (TeV and MI) |
| Optical Transition Radiation Detector | Fast Bunch Integrators (TeV and MI) |
| Sampled Bunch Display (TeV and MI) | Tevatron Abort Gap Integrator |
| Ionization Profile Monitors (TeV and MI) | Beam Current Monitors (TeV, MI and Anti-proton Source) |
| Tevatron Synchrotron Light Detector | Transport Line Beam Position Monitors |
| | Recycler AC Beam Current Monitor |
| | Recycler Transverse Damper |
| | Main Injector Antiproton Damper |
| | Recycler Beam Line Tuner |
| | Booster Damper |

## 2. LabVIEW

LabVIEW is a graphical, data driven programming language developed and supported by National Instruments Corporation. LabVIEW is one of a family of software and hardware products that support instruments and instrumentation systems. Using LabVIEW the Instrumentation Department has been able to integrate with older legacy VME hardware. It can do this due to its support for older drivers, and the use of the MXI PC to VME interface system. This capability has allowed developers to upgrade complex systems while using existing hardware. This also allows LabVIEW to integrate new VME COTS boards into systems for functionality not found in strictly National Instruments hardware.

LabVIEW running on PC hardware has been a good fit for a number of Instrumentation systems. Features in LabVIEW that have been leveraged for efficient implementation of these systems include:

- integration of a variety of legacy hardware using extensive hardware driver support
- ease of integrating C language drivers for custom hardware
- graphical debugging tools
- rapid development and modification cycles
- rich graphical libraries for GUIs

- data/signal analysis support
- portability between OS versions, LabVIEW versions, and PCs

**2.1 Architecture – LabVIEW on PC Hardware**

LabVIEW-based instrumentation systems have been implemented using both Mac™ and Windows™ OS systems. The majority of systems are Windows-based due to the broader National Instruments hardware driver support for Windows.

Much of the Tevatron-era Instrumentation uses legacy VME devices. The VXI-MXI-2 interface from National Instruments allows LabVIEW based systems to make use of VXI as a field bus for most of our systems. To integrate these systems into the accelerator complex, LabVIEW libraries were developed which support communication with the ACNet Control system.

Most Tevatron era instrumentation systems include additional I/O devices. These devices communicate using PCI, USB, and GPIB. GPIB devices are typically oscilloscopes, and power supplies, both low, and high voltage. Drivers for these interfaces employ the Virtual Instrument Software Architecture (VISA) to ensure portability. Instrumentation systems deployed in the last few years have included custom designed PCI hardware. These devices required development of C programming language based Dynamic Linking Libraries (DLLs) that are used via LabVIEW external code support.

All of the instrumentation systems make extensive use of the graphical user interface (GUI) capabilities in LabVIEW. The main instrument applications typically include local control and live status as well as raw and analyzed data displays. These features are primarily used during system development and debugging, after which the systems are controlled remotely through ACNet. The main application GUI remains useful for periodic system performance checks, and debugging system failures.

Since LabVIEW Instruments are located in remote locations around the Tevatron ring, we use Timbuktu™ and Windows Remote Desktop Connection™ for periodic maintenance and debugging tasks. Timbuktu is particularly useful for situations requiring system monitoring by several system experts simultaneously.

**2.2 Example: Tevatron Ionization Profile Monitor**

The TeV Ionization Profile Monitor (IPM) [6] application is typical of the LabVIEW Instruments. The main application displayed in Figure 1 shows various sections that allow setup, control, status and data display. Beyond the typical hardware setup and diagnostic capabilities accessible in the Configure section is the means for setting up measurement specific hardware settings. There are 40 different measurement specific configurations which can be activated via manual control, but which are typically selected automatically via ACNet.

Control functions provided in the Command section allow manual control and intervention in measurements. Since measurements require a specific accelerator clock signal for a trigger, this allows operation to be decoupled from the Tevatron state. Controls in the Log File section allow a user to save an interesting measurement if logging was not requested in the measurement specification, as well as recall previous measurements for re-analysis and display.

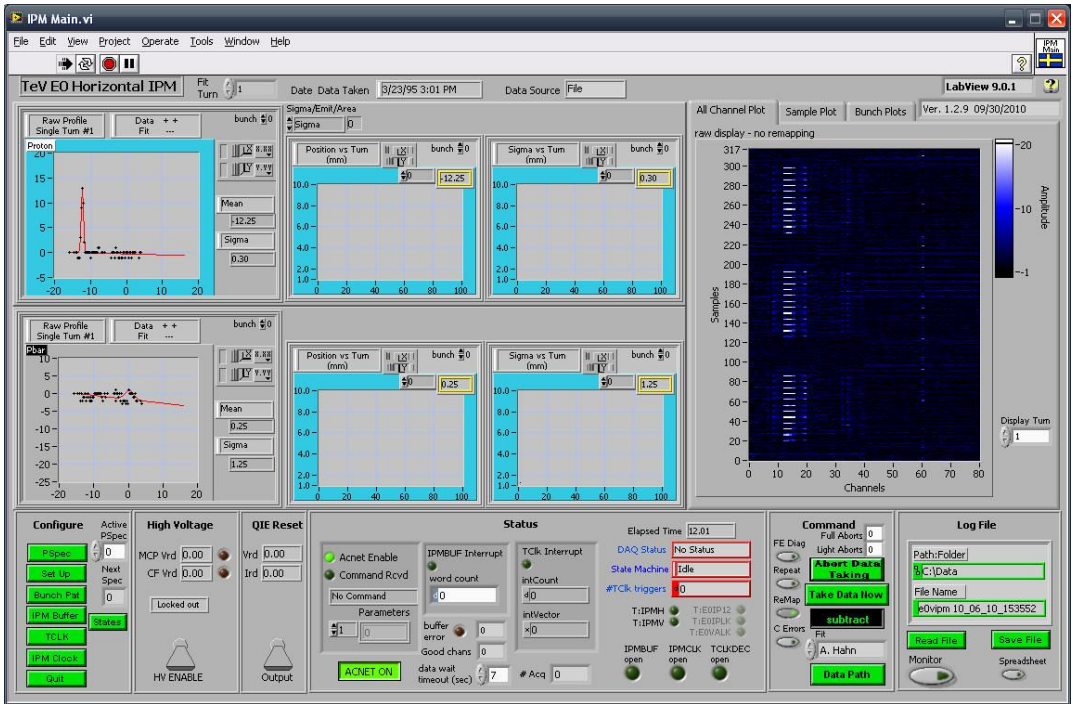

**Figure 1.** LabVIEW main application page for the Tevatron Ion Profile Monitor

The TeV IPM uses custom designed PCI boards for timing and data buffering. The Status section contains status indicators for this hardware, as well as showing the current state of the state machine in the main application. More advanced hardware status is available by selecting the hardware sub-system buttons in the Configure section.

The main application is dominated by analyzed and raw data displays, as these give an immediate indication of system performance. The use of the data displays is determined in part by the measurement specification, so not all displays contain information for each measurement.

## 3. VxWorks on VME

Early Run II instruments like the first generation beam position monitors were built with Intel Multibus hardware, Zilog Z80 microprocessors and were coded in assembly language running in rudimentary homegrown software operating environments. More recently these systems have been updated to the current standard architecture. These systems are built with VME resident hardware, PowerPC based single board computers and software written in C/C++ running on the Wind River VxWorks real-time operating system.

The VxWorks platform allows for fast reading, manipulation and return of data. FPGA technologies and ever faster crate controller boards have allowed huge increases in remote computational power while still returning data to the ACNet system in a timely manner for realtime plotting. Most instrument VxWorks front-ends can sustain 720 Hz realtime data return to the ACNet system's Fast Time Plot (FTP) and Snapshot (SNP) packages. This data can also be captured in the Shot Data Analysis (SDA) system for long term storage and analysis.

VxWorks running on VME allows access to a large variety of COTS boards to use in our systems. Much of the COTS hardware is taken from audio, video and radar applications. These industries use ADC technologies that coincide with much of what needs to be measured in accelerator applications. For example; all the DC Current Transformer front-ends use ADCs developed for audio recording. The Beam Position Monitors use digital down converters developed for both civilian and military radar systems. Use of COTS devices has allowed more rapid development needed for new systems.

**3.1 VME Software Architecture**

Current VME based instruments are implemented with a layered software architecture to provide design clarity and to isolate the instrumentation performance from the effects of the ACNet system operation. Figure 2 depicts these modular and layered software relationships. The three major blocks depicted represent the high level requirements of any instrument front-end; Global Control, Engineering Support and the Instrument blocks

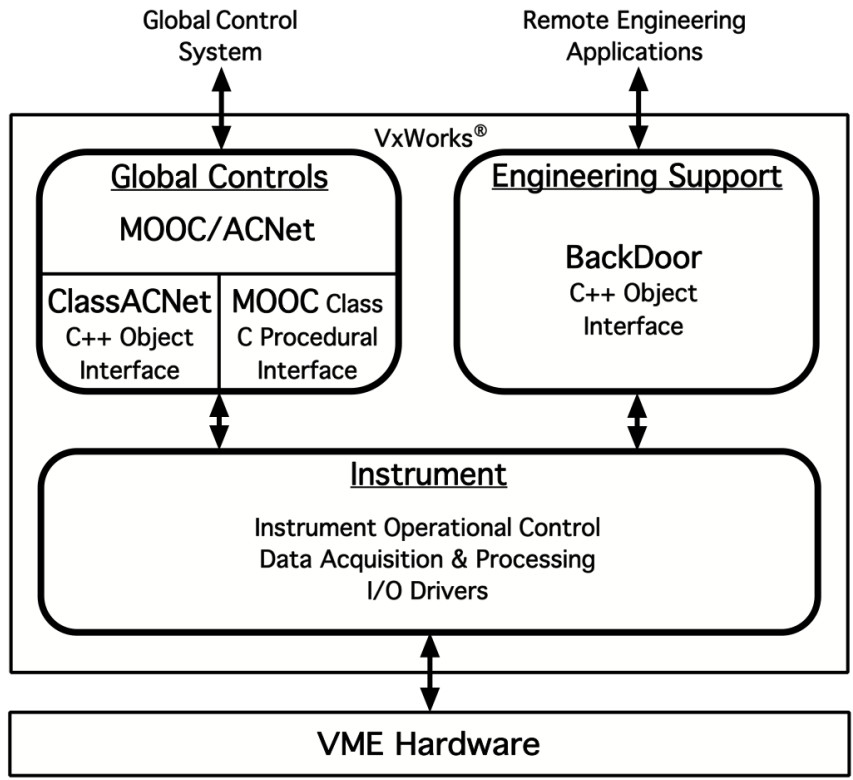

**Figure 2.** The instrumentation modular software design for VxWorks on VME.

**3.1.1 Global Controls Block**

The ACNet system communicates with the instrument front-end using basic reading/setting protocols as provided by the Controls Department. Another Controls Department software package called Minimal Object Oriented Communications (MOOC) adds higher-level protocols for boot-time parameter initialization, alarms FTP and SNP data plotting.

C language programmers connect instrument data and parameters to the MOOC/ACNet software by implementing MOOC callback functions that access the appropriate values. This procedural interface to a given instrument reading is encapsulated into an object like structure called a MOOC class. MOOC classes are often copied and modified as needed for new functionality.

For C++ language programmers the Instrumentation Department created a higher level software layer called ClassACNet designed to encapsulate the MOOC/ACNet software, provide an easy to use object-based interface between instrument data structures and the ACNet system, and to encourage software reusability. ClassACNet defines two classes which form the basis for communicating with the ACNet system: class Portal is layered on top of MOOC/ACNet to provide full-featured control system functionality for all data objects, and class ObjAccessor ties data objects to a given Portal instance. ObjAccessor is a pure virtual class and must be inherited by user classes, called accessors, which handle the details of data acquisition or control. Each Portal instance supports multiple ObjAccessor instances, and instantiating an accessor automatically establishes a data connection to the ACNet system. This provides a plug-in architecture where the programmer attaches instrument data objects to the ACNet system by simply instantiating accessors. The ClassACNet software package contains a collection of off-the-shelf accessors to address most common data and parameter communication needs. Accessors are included for reading and setting memory based data in situ, for instantiating and manipulating scalars, vectors and matrices, for calling user provided C-style callback functions and for invoking the constructor, accessor and mutator methods of other user defined classes. In cases where unique processing is required the user may want to create custom accessors, in which case the ClassACNet library provides a rich set of examples.

### 3.1.2 Engineering Support Block

It is often desirable to access instrument data and internal data structures in a user friendly way for diagnostic and debugging purposes. Some of these data may not be needed in the ACNet system, and perhaps some should remain hidden from the casual user. To meet this need the Instrumentation Department created an out-of-band communication protocol called BackDoor for exchanging data and parameter values between instruments and remote development computers. The BackDoor software architecture is similar to that of ClassACNet having BackDoorServer instances and plug-in BackDoorAccessor instances that make data available to remote clients. For security reasons each server instance maintains an access control list to limit access to specified client IP addresses or address ranges. A BackDoor server has been implemented in C++ for VxWorks and a client has been implemented in LabVIEW. The use of LabVIEW on the client side allows rapid development of sophisticated graphical diagnostic applications with a relatively small amount effort. As with ClassACNet, a package of off-the-shelf accessors is provided to make the instrument programmer's life easier. Standard BackDoor accessors are provided for all of the data types supported by ClassACNet as well as special accessors that support block transfer direct memory access to the standard VME memory address spaces. In addition, a collection of diagnostic accessors provides tools for measuring and displaying the real-time performance of the instrument code. There are accessors for histogramming the execution time of code segments and repetitive code periods, collecting and displaying code execution time lines, and even a software logic analyzer with programmable triggering.

### 3.1.3 Instrument Block

The instrument block is where the software meets the hardware to coordinate and control the instrument's base functionality. As the figure implies software layering is applied in this block to separate input/output activity from data acquisition and operational control. The VxWorks operating system provides a powerful platform for hosting instrument software development. In some cases commercial hardware vendors provide off-the-shelf I/O drivers targeted to the operating system. In other cases, particularly when the hardware is developed in-house, the I/O drivers are coded as task level driver packages to simplify the design and improve performance.

## 4. Conclusion

LabVIEW and VxWorks give the Instrumentation Department a variety of tools that have been used to create the unique solutions for the Tevatron's Run II environment. Both have been needed for their different strengths. Both have provided good solutions that have aided in the Run II successes. Both will continue to be used for future programs at Fermilab.

## Acknowledgments

We would like to thank all those who have made instrument front-ends in the Run II era. Alan Baumbaugh, Willem Blokland, Charles Briegel, Robert Flora, Ken Fullett, Steve Foulkes, Alan Hahn, Sharon Lackey, Eugene Lorman, Dennis Nicklaus, Luciano Piccoli, Randy Thurman-Keup, Jeff Utterback, James Zagel and Dehong Zhang. Fermilab is operated by Fermi Research Alliance, LLC under Contract No. DE-AC02-07CH11359 with the United States Department of Energy.